\documentstyle[12pt,aps,epsfig]{revtex}
\newcommand{\be}{\begin{equation}}
\newcommand{\ee}{\end{equation}}
\newcommand{\bea}{\begin{eqnarray}}
\newcommand{\eea}{\end{eqnarray}}

\def\Journal#1#2#3#4{{#1} {#2} (#4) #3}

\def\NP{{Nucl. Phys.}}
\def\PL{{Phys. Lett.}}
\def\PR{{Phys. Rep.}}
\def\PRL{Phys. Rev. Lett.}
\def\PRD{{Phys. Rev.} D}
\def\PRC{{Phys. Rev.} C}

\def\ZPC{{Z. Phys.} C}
\def\JPG{{J. Phys.} G}

\def\AP{{Ann. Phys. (N.Y.)}}

\def\CPC{{Comput. Phys. Comm.}}
\def\RMP{{Rev. Mod. Phys.}}

\begin{document}
\title{Inverse Slope Systematics in High-Energy p+p
and Au+Au Reactions\footnote[2]{Supported by AvH Foundation and DAAD.}}
\author{A. Dumitru}
\address{Physics Department, Yale University\\
P.O.\ Box 208124, New Haven, CT 06520, USA\\}
\author{C. Spieles}
\address{Nuclear Science Division, Lawrence Berkeley National Laboratory\\
Berkeley, CA 94720, USA}
\date{\today}
\maketitle   
\begin{abstract}
We employ the Monte-Carlo PYTHIA to calculate the transverse mass spectra
of various hadrons and their inverse slopes $T^*$ at $m_T-m=1.5-2$~GeV in
p+p reactions at $\sqrt{s}=200$~GeV. Due to (multiple) minijet production
$T^*$ in general increases as a function of the hadron mass. Moreover, the
$T^*(m)$ systematics has a ``discontinuity'' at the charm threshold, i.e.\ the
inverse slope of $D$-mesons is much higher than that of non-charmed hadrons
and even of the heavier $\Lambda_C$ baryon.
The experimental observation of this characteristic behaviour in 
Au+Au collisions would indicate the absence of c-quark rescattering.
In contrast, the assumption of thermalized partons and hydrodynamical
evolution would lead to a smoothly increasing $T^*(m)$,
without discontinuity at the charm threshold.
The degree of collective transverse flow, indicated by
the slope of the $T^*(m)$ systematics, depends strongly on whether 
kinetic equilibrium is maintained for some time after hadronization or not.
\end{abstract}
\newpage

Experimental data on single-inclusive hadron production in p+p reactions at
$\sqrt{s}=23-63$~GeV show nearly exponential transverse mass spectra at
low transverse momentum~\cite{Alper}.
Moreover, the inverse slopes (``apparent temperatures'')  are practically
the same for pions, kaons, protons and their antiparticles, i.e.\
they do not depend on the hadron mass.
The observed deviation from this behaviour in nucleus-nucleus collisions has 
been interpreted as a signature for collective transverse flow~\cite{Bea97}.

In this letter, we discuss how the inverse slope systematics extends to
higher energies, i.e.\ p+p reactions at $\sqrt{s}=200$~GeV, and
transverse momenta on the order of a few GeV. In this kinematic domain,
minijet production and fragmentation gives an important contribution.

We shall also discuss Au+Au collisions, where minijets might rescatter
substantially. This, in turn, should reflect in
a characteristic mass dependence of the inverse slopes. We will confront
the predictions of two extreme scenarios: superposition of minijet
production (and fragmentation) without final-state interactions (as in p+p
reactions) versus local thermalization of parton matter undergoing
hydrodynamical expansion.

In p+p reactions at $\sqrt{s}=200$~GeV,
we expect that the hadrons with transverse masses on the order of a few GeV
are dominantly produced via fragmentation of minijets. As a first step,
we estimate the transverse momentum distribution of
$c$-quarks at midrapidity employing the well-known expression for
inclusive single-jet production within perturbative QCD (pQCD) in
leading-logarithm approximation (LLA)~\cite{Owens},
\be \label{sngljet}
E\frac{d^3 \sigma}{d^3p}\left(pp\rightarrow c\overline{c}+X\right) = 
\int dx_a dx_b\, G\left(x_a,\mu^2\right) 
G\left(x_b,\mu^2\right) \frac{\hat{s}}{\pi}
  \frac{d\sigma^{gg\rightarrow c\overline{c}}}{dt} 
  \delta\left(\hat{s}+\hat{t}+
  \hat{u}-2m_c^2\right)\quad.
\ee
$\hat{s}$, $\hat{t}$, $\hat{u}$ are the usual Mandelstam variables of the
parton-parton scattering subprocess, and $d\sigma^{ab\rightarrow cd}/dt$
denotes its differential cross section in lowest order of perturbative QCD.
For simplicity, we take into account only the contribution from the
$gg\rightarrow q\overline{q}$ process. This is sufficient to illustrate our
point.

The expression for $d\sigma/dt$ that
accounts for the finite quark-mass is rather lengthy and can be found in
the literature, cf.\ e.g.~\cite{ggQQ}. We therefore omit it here.
We assume $m_c=1.5$~GeV, $\Lambda_{QCD}=300$~MeV, and
evaluate the strong coupling constant at the momentum transfer scale
given by the transverse mass of the produced quark,
$Q^2=m_T^2=m_c^2+p_T^2$~\cite{VZPC}.

$G(x,\mu^2)$ denotes the LO gluon distribution function
in the proton, which we take from ref.~\cite{GRV95}. Since we work only
in LLA, we employ the same
momentum scale in the parton distribution functions as in the strong
coupling constant, i.e.\ $\mu^2\equiv Q^2$.

In principle, the hadron spectra could be calculated by convoluting
the expression for jet production with fragmentation functions~\cite{Owens}.
However, for transverse masses of a few GeV such an analysis would at best
be qualitative since, e.g., multi jet production and initial state radiation
are not included. Also, in this domain the fragmentation functions suffer
from logarithmic infrared divergences which have to be regulated by a model
for soft particle production.

Therefore, to calculate the hadron transverse mass spectra we rather employ the
PYTHIA event generator~\cite{Sjo94}, using the default parameter
settings (version~6.115). PYTHIA simulates high energy hadronic and 
leptonic interactions by implementing a
large number of hard and soft (sub-)processes, and in particular, a scheme
for the nonperturbative hadronization mechanism. 
The model goes significantly beyond pQCD in LLA. 
It describes not only single-inclusive
minijet production but also includes multi-jet production and initial state
radiation. PYTHIA is designed to model the complete event structure 
(like jet profiles, multiplicity fluctuations, various types of 
correlations etc.) and it has been shown to agree reasonably well with 
experimental observations at collider energies~\cite{SjoZ}.

The Lund string scheme~\cite{And83} is an integral part of the model used to 
describe the fragmentation of (mini-)jets. They are modeled as one dimensional 
color flux tubes which decay into hadrons: quark-antiquark
pairs tunnel in the color field and the field energy is transformed into the
sum of the 
transverse masses $m_T$\footnote{Note that 'transverse' is defined with 
respect to the string axis. Nothing is said here about the orientation 
of the string with respect to the beam axis.}. 

The tunnel probability is proportional to $\exp(-{\pi m_T^2}/{\kappa})$,
where $\kappa$ is the string tension. Thus, the creation of quarks 
with high transverse momentum is heavily suppressed. As a consequence, also
the produced hadrons cannot acquire large values of $m_T$.
The above formula also leads to a strong suppression of heavy quarks. The
probability for producing a light quark as compared to a charm quark is
about $1:10^{-11}$~\cite{And83}.
The energy $E$ and the longitudinal momentum $p_z$ of the produced hadrons are
determined by an iterative scheme: for each hadron the fragmentation
function $f(z)$ determines the probability that the hadron picks 
a fraction $z$ out of the available $E+p_z$. The default
fragmentation function used in PYTHIA reads
$f(z)\sim z^{-1}(1-z)^{0.3} \exp (-0.58 {\rm GeV^{-2}} m_T^2/z)$.

The Lund string model has been shown to succesfully describe
the nonperturbative hadronization in $e^+e^-$ annihilation
events~\cite{And83}. Moreover,
the concept of a color flux tube, fragmenting according to a universal
fragmentation scheme, has been carried over to hadron-hadron
interactions. The microscopic models FRITIOF, RQMD, and UrQMD~\cite{Micro}
utilize string fragmentation routines for the simulation 
of soft particle production in p+p, p+A and A+A reactions.
The only difference to strings from $e^+e^-$ annihilation
are the leading valence (di-)quarks --- the remnants of the incident 
hadrons --- as string end-points. The excitation of the strings 
is due to single or double diffractive interactions which can be understood
and parametrized in the framework of Regge theory (see e.g.~\cite{Sch94}).
These nonperturbative processes account for the major part of the total 
cross sections at CERN-SPS energies or higher, $\sqrt{s}> 20$~GeV. 
Strings which are excited in these processes are preferentially oriented
along the beam axis of the incident hadrons, leading to much higher
typical longitudinal than transverse momenta of produced hadrons.

\begin{figure}[htp]
\centerline{\hbox{\epsfig{figure=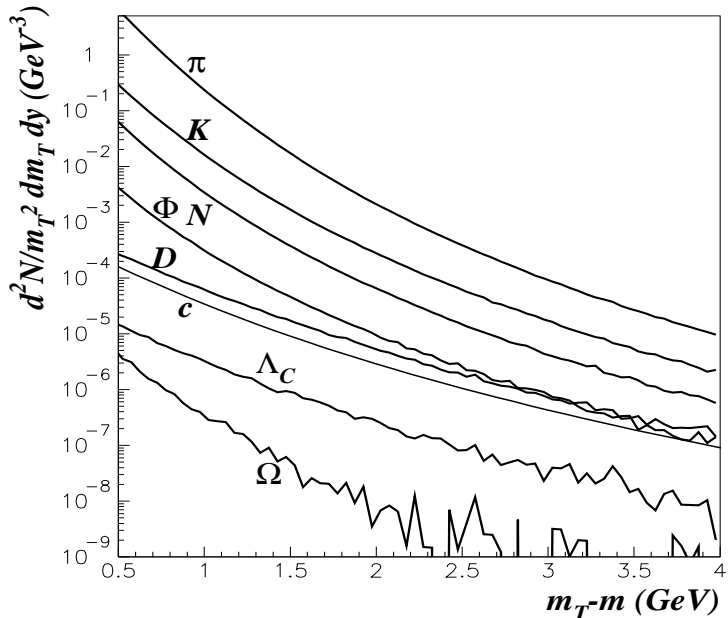,height=9cm,width=10cm}}}
\caption{Transverse mass spectra (at midrapidity, $y=0$) of various hadrons
in p+p reactions at $\protect\sqrt{s}=200$~GeV, as calculated 
with PYTHIA~6.115. The spectra of the individual hadron species include all 
isospin projections and charge conjugated states.
The c-quark spectrum (without $\overline{c}$-quarks) is calculated within
pQCD in LLA.}
\label{dndmt}
\end{figure}  
In the kinematic region where perturbative minijet production becomes
important, the color flux tubes may no longer be oriented longitudinally,
but according to the pQCD subprocess that produces the minijets acquire
significant transverse momentum. In the frame were the $z$-axis
is parallel to the flux tube, the hadrons are still produced
according to the above fragmentation function. 
However, the string axis and the particles' momenta are now rotated
with respect to the lab frame.

Figure~\ref{dndmt} depicts the resulting transverse mass spectra. One observes
that the PYTHIA-spectra follow nonexponential distributions, remnant of the
perturbative QCD processes that describe the minijet production. Also,
the ``stiffness'' of the spectra increases with the mass of the hadron. This
will be discussed in more detail below. In particular, the slope of the
$D$-meson spectrum equals that of the
$c$-quarks, which are produced purely
by perturbative parton-parton scattering\footnote{To obtain the number
distribution of $c$-quarks we have simply divided the differential
cross section, eq.~(\ref{sngljet}), by 40~mb. If we multiplied by two
(to include also $\overline{c}$-quarks) the quark and $D$-meson
spectra would coincide.}.
The reason is that $D$-mesons can only be
produced as the leading hadron from a $c/\overline{c}$ quark jet since
the tunneling probability of a $c-\overline{c}$ pair in a color flux tube
is practically zero (as discussed above).
The $m_T$-distribution of $\Lambda_C$-baryons, on the
other hand, is slightly ``softer'' since
it involves tunneling of a diquark-antidiquark pair out of the
vacuum (besides the perturbative production of a $c$-quark) .

\begin{figure}[htp]
\centerline{\hbox{\epsfig{figure=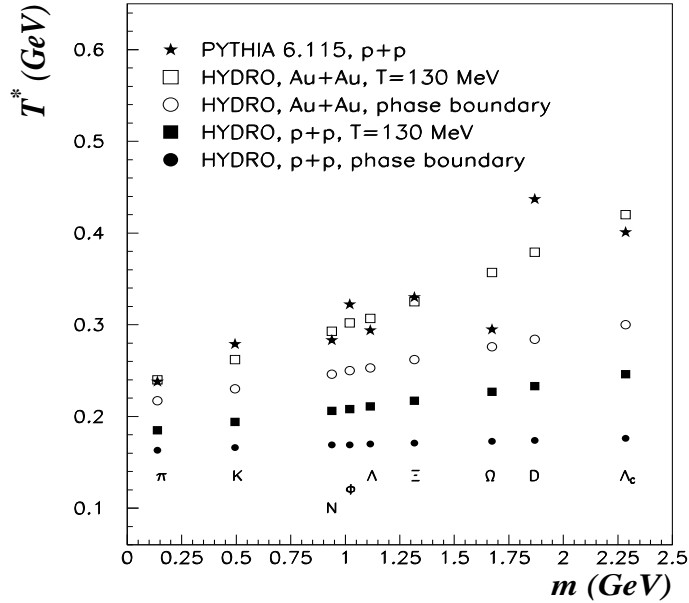,height=9cm,width=10cm}}}
\caption{Inverse slopes (at midrapidity, $y=0$) as a function of hadron mass;
PYTHIA~6.115 predictions for p+p at $\protect\sqrt{s}=200$ GeV, and 
results from hydrodynamics of p+p and Au+Au (calculated on the boundary
between mixed and hadronic phase and on the $T=130$~MeV isotherm,
respectively).}
\label{tm}
\end{figure}

In Fig.~\ref{tm}, we show the inverse slopes as a function of hadron mass.
We compute the inverse slope by a fit of the transverse
mass spectrum to a Boltzmann distribution,
\be
\frac{1}{m_T^2}\frac{d^2N}{dm_T dy} \propto \exp\left(
 -m_T/T^*\right)\quad,
\ee
We restrict the fit to the range $m_T-m\in\left[1.5,2\right]$~GeV.

As already mentioned in the introduction, in p+p reactions at lower energies,
the apparent temperatures at small $p_T$ were found to
be about the same for pions, kaons, protons and their
antiparticles~\cite{Alper,Bea97}.
This changes at higher $p_T$ and $\sqrt{s}$ due to the contribution from
minijets.
The single inclusive cross section at midrapidity
and $m_T-m\ge1$~GeV is
dominated by color flux tubes that are no longer oriented longitudinally,
but have significant transverse momentum (due to the pQCD subprocess).
This leads to much
higher $m_T$ than in case of longitudinally oriented strings. Moreover,
the inverse slopes of the $m_T-m$ distributions in the lab frame are
strongly $m$-dependent.
Such an increase of the inverse slope with particle mass
can also be extracted from ${\rm p +\bar p}$ collider 
experiments~\cite{UAexp} at $\sqrt{s}=540$~GeV, and from the parametrization
of the $p_T$-distributions given in ref.~\cite{Bourquin}.
Although that parametrization was restricted to the energy region
$\sqrt{s}<63$~GeV, it yields slopes for the pi-, phi-, and D-mesons
that agree with those obtained from PYTHIA
to within $20\%$, if one simply extrapolates it to $\sqrt{s}=200$~GeV.

The qualitative reason for this behaviour can be traced back to 
$e^+e^-$ annihilation processes which can be modeled as the 
fragmentation of a string with fixed energy. The observed kinetic energy 
distribution of produced hadrons show a considerably harder spectrum 
for kaons and protons than for pions~\cite{epluseminus}.

Thinking of the initial state in A+A collisions as a
superposition of p+p reactions one could
attribute any change of the particle slopes to final state interactions.
The multiple soft $p_T$-kicks that a projectile nucleon can experience
while propagating through the target, i.e.\ the Cronin effect,
has been shown to be small for pions produced in Au+Au at
$\sqrt{s}=200A$~GeV~\cite{Gyu98}.
However, this does not automatically hold true for heavier hadrons.
Since we have not made any attempt to include such multiple scattering
effects, nor modifications of the parton distribution functions in
nuclei, we restrict the application of the minijet/string fragmentation model
to p+p reactions.

In ultrarelativistic head-on collisions of heavy nuclei ($A\sim 200$),
the situation might, however, be very different as compared to the p+p case.
Transport calculations~\cite{Geiger} suggest that the initially directed
momenta of the partons could be quickly redistributed through rescattering.
This would then lead to the formation of a very hot ($T\approx300$~GeV)
vacuum of ``macroscopic'' size (volume $\sim100$~fm$^3$),
offering the opportunity to study hot QCD.

Indeed, if minijets with transverse masses of a few GeV thermalize quickly,
energy densities in excess of 10~GeV/fm$^3$ can be reached~\cite{Geiger}
(for Au+Au at $\sqrt{s}=200A$~GeV). 
This energy density is much higher than
in p+p since the contribution due to
minijets increases as $A^{2/3}$~\cite{KLL}.

For definitness we consider Au+Au collisions at $\sqrt{s}=200A$~GeV that will
be studied at the Relativistic Heavy Ion Collider (RHIC) in Brookhaven in the
near future. As the extreme case we assume that the produced minijets
rescatter so frequently that they thermalize locally. The subsequent evolution
is then described within hydrodynamics.

The pQCD processes that produce the minijets in the central region occur on
a time scale of $1/m_T\simeq0.1$ fm. We assume that one to two rescatterings
per
particle are necessary for local thermalization, and take $\tau_0=0.6$ fm.
This seems also reasonable in view of the fact that with $\tau_0=1$ fm one
is able to reproduce the measured single particle spectra for central Pb+Pb
reactions at $\sqrt{s}=18A$ GeV, cf.\ e.g.\ \cite{DumRi}. At the higher
center of mass energy of RHIC, the parton density in the central region
increases, and therefore a smaller thermalization time is expected, cf.\
also \cite{Geiger,therm}.

As initial conditions we assume a net baryon rapidity density of $dN_B/dy=25$,
and an energy density of $\epsilon_0=17$~GeV/fm$^3$.
Employing the formula of Bjorken~\cite{Bj}, $\epsilon_0\tau_0\pi R_T^2=
dE_T(\tau_0)/{dy}$,
with a nuclear radius of $R_T=6$~fm, we obtain an initial transverse energy
at midrapidity of $dE_T/dy=1.2$~TeV.
We have extracted this value for $dE_T/dy$ for central ($b<2$~fm) Au+Au
collisions at $\sqrt{s}=200A$~GeV from the minijet/string
fragmentation model FRITIOF~7.02~\cite{FRIT}. It is also compatible with the
prediction of HIJING~\cite{Gypc}. Note that for an isentropic hydrodynamical
expansion $dE_T/dy$ decreases with time. On the hadronization hypersurface,
we obtain $dE_T/dy=640$~GeV~\cite{DumRi}.

The initial (net) baryon density at
midrapidity, $\rho_0$, is given by a similar expression as $\epsilon_0$
above, except that $dE_T/dy$ is replaced by $dN_B/dy$.
These densities are (initially) assumed to be distributed in the transverse
plane according to a so-called ``wounded nucleon'' distribution,
$\epsilon(\tau_i) = \epsilon_0 f(r_T)$, $\rho(\tau_i) = \rho_0 f(r_T)$,
with $f(r_T) = \frac{3}{2}\sqrt{1-r_T^2/R_T^2}$.

In order to respect boost-invariance, we require the longitudinal flow to
have a ``scaling flow'' profile, $v_z=z/t$~\cite{Bj,KMcL}. Cylindrically
symmetric transverse expansion~\cite{DumRi,Baym,Kat86} is superimposed.
For $T>T_C=160$~MeV we employ the well-known MIT bagmodel equation of state,
$p=(\epsilon-4B)/3$, where $p$ denotes the pressure and $B$ the energy density
of the QGP at $T=0$ and vanishing net baryon charge.
For simplicity we assume an ideal gas of quarks,
antiquarks (with masses $m_u=m_d=0$, $m_s=150$~MeV), and gluons.

For $T<T_C$ we assume an ideal hadron gas that includes the complete
hadronic spectrum up to a mass of 2~GeV.
At $T=T_C$ we require that both pressures are equal, which
fixes the bag constant to
$B=380$~MeV/fm$^3$. The normalization is such that for $T\rightarrow0$
the pressure of the nonperturbative vacuum (i.e.\ that of the hadronic phase)
vanishes.
By construction the EoS exhibits a first-order phase transition. This
``softening'' of the EoS in the transition region strongly reduces the
tendency of matter to expand on account of its pressure~\cite{Kat86,soft}.
For a more detailed
discussion of the initial conditions and expansion dynamics in Au+Au at RHIC
energy please refer to ref.~\cite{DumRi}.

For comparison, we have also extracted the inverse slopes from hydrodynamics
in p+p at $\sqrt{s}=200$ GeV. In this case, we employ $R_T=1.18$ fm,
$f(r_T)=\Theta(R_T-r_T)$, $dN_B/dy=0$, $dE_T/dy=2.8$ GeV (as obtained from
PYTHIA), and (for simplicity) the same $\tau_0$ as for Au+Au.

In Fig.~\ref{tm} we compare the PYTHIA
predictions for p+p with those of hydrodynamics of p+p and Au+Au.
Within the hydrodynamical solution for Au+Au, strong
collective flow of quark-gluon matter\footnote{The average flow velocity on the
phase boundary to purely hadronic matter is approximately one third of the
velocity of light~\cite{DumRi}.} Doppler-shifts $T^*$ far above the real
emission temperature $T_C$. If kinetic equilibrium in the hot hadron gas
is maintained for some time after hadronization (say, until the temperature
drops to 130~MeV), the collective transverse flow can increase even further,
and $T^*\sim400$~MeV can be reached for the charmed hadrons
(cf.\ open squares in
fig.~\ref{tm}). Hadrons produced in the expanding QGP
can thus reach comparably ``stiff'' $m_T$-spectra as those produced in p+p
via minijets.

One also observes that in hydrodynamics $T^*$ is nearly proportional to $m$.
In particular, there is no jump in $T^*$ at the charm threshold and the
inverse slope of the $\Lambda_C$ is larger than that of the $D$, unlike
in the minijet/string fragmentation model. In a thermal environment (without
interactions), the mass is the
only hadron-specific quantity that enters its momentum distribution. Thus,
if the perturbatively produced $c-\overline{c}$ pairs equilibrate with the
QGP, the inverse slope of the $D$-mesons is significantly
smaller than in p+p reactions at the same energy per nucleon (cf.\ also
the discussion of $\langle p_T\rangle_D$ in refs.~\cite{DumRi,SvU},
and~\cite{cdEdx} for the effect of $c$-quark energy loss
on lepton radiation in ultrarelativistic heavy-ion collisions).

In p+p reactions, on the other hand, the initial energy density $\epsilon_0
=1.1$ GeV/fm$^3$ is much smaller than in Au+Au. In fact, for our choice of
initial conditions the initial state is not in the pure QGP phase but in the
phase coexistence region. Consequently, on the hadronization hypersurface
there is practically no collective transverse flow, and the inverse slopes
of the various hadrons are similar, and equal to the real emission temperature
$T=T_C$. If freeze-out occurs deeper in the hadronic phase, e.g.\ on the
$T=130$ MeV isotherm, a small flow is created due to rescattering in the
purely hadronic phase. The pressure in p+p reactions at $\sqrt{s}$ of a few
hundred GeV can not exceed that of the nonperturbative vacuum by far, and we
thus find no significant transverse expansion~\cite{SZ}. Therefore,
in contrast to the minijet/string fragmentation model hydrodynamics can not
reproduce the experimentally observed~\cite{UAexp} bending of the $m_T$
distributions and the increase of $T^*$ with $m$.
 
In summary, we have shown that in p+p reactions at high energy the
inverse slopes $T^*$ of the transverse mass spectra (at midrapidity and
$m_T-m$ of a few GeV)
of various hadrons are correlated to their mass. This is due to the
underlying pQCD subprocess (i.e.\ minijet production) that produces
fragmenting color strings that are not parallel to the beam axis.
$T^*(m)$ shows a very
strong increase at the charm threshold, i.e.\ the inverse slope of the
$D$-mesons is much higher than that of the $\Omega$-baryons.

If a dense QGP is created in central Au+Au collisions, in which 
$u$-,$d$-,$s$-, and $c$-quarks,
and the gluons (up to a few GeV of $p_T$) equilibrate kinetically, the
inverse slope of the $D$-mesons
decreases substantially as compared to the p+p case, and that
of the $\Omega$-baryons increases. Collective transverse flow of such a
hypothetical quark-gluon fluid would establish a nearly linear
relationship between the inverse slopes and the hadron masses.
Thus, the $T^*(m)$ systematics at $m_T-m\simeq1-3$~GeV provides an opportunity
to experimentally determine the degree of heavy quark
rescattering and equilibration in relativistic heavy ion collisions.\\
{\bf Acknowledgements:}
We thank M.\ Gyulassy, B.\ M\"uller, and T.\ Ullrich for helpful discussions.
A.D.\ gratefully acknowledges a postdoctoral fellowship by the German
Academic Exchange Service (DAAD).
C.S.\ thanks the Nuclear Theory Group at the LBNL for support and kind
hospitality. C.S.\ is supported by the Alexander von Humboldt Foundation
through a Feodor Lynen Fellowship.


\begin{references}
\bibitem{Alper} B. Alper et al., \Journal{\NP}{B100}{237}{1975};
K. Guettler et al., \Journal{\NP}{B116}{77}{1976}
\bibitem{Bea97} I.~G.~Bearden et al. (NA44 Collaboration),
\Journal{\PRL}{78}{2080}{1997}
\bibitem{Owens} J.F. Owens, \Journal{\RMP}{59}{465}{1987}
\bibitem{ggQQ} M. Gl\"uck, J.F. Owens, E. Reya, \Journal{\PRD}{17}{2324}{1978}
\bibitem{VZPC} R. Vogt, \Journal{\ZPC}{71}{475}{1996}
\bibitem{GRV95} M. Gl\"uck, E. Reya, A. Vogt, \Journal{\ZPC}{67}{433}{1995}
\bibitem{Sjo94} T.~Sj\"ostrand, \Journal{\CPC}{82}{74}{1994}
\bibitem{SjoZ} T.~Sj\"ostrand, M. van Zijl, \Journal{\PRD}{36}{2019}{1987}
\bibitem{And83} B.~Andersson, G.~Gustafson, G.~Ingelman, T.~Sj\"ostrand,
                \Journal{\PR}{97}{31}{1983}
\bibitem{Micro} B. Nilsson-Almqvist, E. Stenlund,
                \Journal{\CPC}{43}{387}{1987};
G. Gustafson, \Journal{\NP}{A566}{233c}{1994};
H.~Sorge, H.~St\"ocker, W.~Greiner, \Journal{\AP}{192}{266}{1989};
S.A. Bass et al., \Journal{Prog. Part. Nucl. Phys.}{41}{225}{1998}
\bibitem{Sch94} G.~A.~Schuler, T.~Sj\"ostrand, \Journal{\PRD}{49}{2257}{1994}
\bibitem{UAexp} 
G. Arnison et al.\ (UA1 Collaboration),
\Journal{\PL}{118B}{167}{1982}; 
M. Banner et al.\ (UA2 Collaboration),
\Journal{\PL}{122B}{322}{1983};
G.J. Alner et al.\ (UA5 Collaboration), \Journal{\PR}{154}{247}{1987}
\bibitem{Bourquin} M. Bourquin, J.-M. Gaillard,
\Journal{\NP}{B114}{334}{1976}
\bibitem{epluseminus} H.~Albrecht et al.\ (ARGUS Collab.),
\Journal{\ZPC}{44}{547}{1989}
\bibitem{Gyu98} M.~Gyulassy, P.~Levai, hep-ph/9807247, hep-ph/9809314
\bibitem{Geiger}
K. Geiger, B. M\"uller,  \Journal{\NP}{B369}{600}{1992};
K. Geiger, \Journal{\PRD}{46}{4965}{1992};
K. Geiger, J. Kapusta, \Journal{\PRD}{47}{4905}{1993}
\bibitem{KLL} K. Kajantie, P.V. Landshoff, J. Lindfors,
\Journal{\PRL}{59}{2527}{1987};
J.P. Blaizot, A.H. Mueller, \Journal{\NP}{B289}{847}{1987};
K. Eskola, K. Kajantie, J. Lindfors, \Journal{\NP}{B323}{37}{1989};
X.N. Wang, M. Gyulassy, \Journal{\PRD}{44}{3501}{1991};
\Journal{\PRD}{45}{844}{1992}
\bibitem{DumRi} A. Dumitru, D.H. Rischke, nucl-th/9806003
\bibitem{therm}
B. M\"uller, X.N. Wang, \Journal{\PRL}{68}{2437}{1992};
E. Shuryak, {\sl ibid} p.\ 3270;
E. Shuryak, L. Xiong, \Journal{\prl}{70}{2241}{1993};
T.S. Biro, E. van Doorn, B. M\"uller, M.H. Thoma,
X.N. Wang, \Journal{\PRC}{48}{1275}{1993};
B. K\"ampfer, O.P. Pavlenko, \Journal{\ZPC}{62}{491}{1994};
K.J. Eskola, X.N. Wang, \Journal{\PRD}{49}{1284}{1994}
\bibitem{Bj} J.D. Bjorken, \Journal{\PRD}{27}{140}{1983}
\bibitem{FRIT}  H. Pi, \Journal{\CPC}{71}{173}{1992};
B. Andersson, G. Gustafson, and H. Pi, \Journal{\ZPC}{57}{485}{1993}
\bibitem{Gypc} M. Gyulassy, private communication
\bibitem{KMcL} K. Kajantie, L. McLerran, \Journal{\NP}{B214}{261}{1983};
K. Kajantie, R. Raitio, P.V. Ruuskanen,  \Journal{\NP}{B222}{152}{1983}
\bibitem{Baym} G. Baym, B.L. Friman, J.P. Blaizot, M. Soyeur, W. Czyz,
\Journal{\NP}{A407}{541}{1983}
\bibitem{Kat86}
M. Kataja, P.V. Ruuskanen, L.D. McLerran, H. von Gersdorff,
\Journal{\PRD}{34}{2755}{1986}
\bibitem{soft} 
C.M. Hung, E. Shuryak, \Journal{\PRL}{75}{4003}{1995};
D.H. Rischke, M. Gyulassy, \Journal{\NP}{A597}{701}{1996}
\bibitem{SvU} B. Svetitsky, A. Uziel, \Journal{\PRD}{55}{2616}{1997};
B. K\"ampfer, O.P. Pavlenko, A. Peshier, M. Hentschel, G. Soff,
\Journal{\JPG}{23}{2001}{1997}
\bibitem{cdEdx} E. Shuryak, \Journal{\PRC}{55}{961}{1997}
\bibitem{SZ} E.V. Shuryak, O.V. Zhirov,  \Journal{\PL}{89B}{253}{1980}
\end{references}
\end{document}